\documentclass[plb,twocolumn,superscriptaddress,amssymb,amsmath,amsfonts,aps]{revtex4}

\preprint{APS/hep-ex}

\usepackage{graphicx}
\usepackage{dcolumn}
\usepackage{bm}

\begin{document}


\title{ A study of the reaction $\pi^-p \rightarrow \omega\pi^- p$ at $18$ GeV/$c$: \\The $D$ and $S$ decay amplitudes for $b_1(1235) \rightarrow \omega\pi$ }

\author{M.~\surname{Nozar}}
\email[]{nozarm@jlab.org}
\altaffiliation[Present address: ]{Thomas Jefferson National Accelerator Facility, Newport News, Virginia 23606}
\affiliation{Department of Physics, Rensselaer Polytechnic Institute, Troy, New York 12180}
\author{G.~S.~\surname{Adams}}
\affiliation{Department of Physics, Rensselaer Polytechnic Institute, Troy, New York 12180}
\author{T.~\surname{Adams}}
\altaffiliation[Present address: ]{Department of Physics, Florida State University, Tallahassee, FL 32306}
\affiliation{Department of Physics, University of Notre Dame, Notre Dame, Indiana 46556}
\author{Z.~\surname{Bar-Yam}}
\affiliation{Department of Physics, University of Massachusetts Dartmouth, North Dartmouth, Massachusetts 02747}
\author{J.~M.~\surname{Bishop}}
\affiliation{Department of Physics, University of Notre Dame, Notre Dame, Indiana 46556}
\author{V.~A.~\surname{Bodyagin}}
\affiliation{Nuclear Physics Institute, Moscow State University, Moscow, Russian Federation 119899}
\author{D.~S.~\surname{Brown}}
\altaffiliation[Present address: ]{Department of Physics, University of Maryland, College Park, MD 20742}
\affiliation{Department of Physics, Northwestern University, Evanston, Illinois 60208}
\author{N.~M.~\surname{Cason}}
\affiliation{Department of Physics, University of Notre Dame, Notre Dame, Indiana 46556}
\author{S.~U.~\surname{Chung}}
\affiliation{Physics Department, Brookhaven National Laboratory, Upton, New York 11973}
\author{J.~P.~\surname{Cummings}}
\affiliation{Department of Physics, Rensselaer Polytechnic Institute, Troy, New York 12180}
\author{K.~\surname{Danyo}}
\affiliation{Physics Department, Brookhaven National Laboratory, Upton, New York 11973}
\author{A.~I.~\surname{Demianov}}
\affiliation{Nuclear Physics Institute, Moscow State University, Moscow, Russian Federation 119899}
\author{S.~P.~\surname{Denisov}}
\affiliation{Institute for High Energy Physics, Protvino, Russian Federation 142284}
\author{V.~\surname{Dorofeev}}
\affiliation{Institute for High Energy Physics, Protvino, Russian Federation 142284}
\author{J.~P.~\surname{Dowd}}
\affiliation{Department of Physics, University of Massachusetts Dartmouth, North Dartmouth, Massachusetts 02747}
\author{P.~\surname{Eugenio}}
\altaffiliation[Present address: ]{Department of Physics, Florida State University, Tallahassee, FL 32306}
\affiliation{Department of Physics, University of Massachusetts Dartmouth, North Dartmouth, Massachusetts 02747}
\author{X.~L.~\surname{Fan}}
\affiliation{Department of Physics, Northwestern University, Evanston, Illinois 60208}
\author{A.~M.~\surname{Gribushin}}
\affiliation{Nuclear Physics Institute, Moscow State University, Moscow, Russian Federation 119899}
\author{R.~W.~\surname{Hackenburg}}
\affiliation{Physics Department, Brookhaven National Laboratory, Upton, New York 11973}
\author{M.~\surname{Hayek}}
\altaffiliation[Permanent address: ]{Rafael, Haifa, Israel}
\affiliation{Department of Physics, University of Massachusetts Dartmouth, North Dartmouth, Massachusetts 02747}
\author{J.~\surname{Hu}}
\altaffiliation[Present address: ]{TRIUMF, Vancouver, B.C., V6T 2A3, Canada}
\affiliation{Department of Physics, Rensselaer Polytechnic Institute, Troy, New York 12180}
\author{E.~I.~\surname{Ivanov}}
\altaffiliation[Present address: ]{Department of Physics, Idaho State University, Pocatello, ID 83209}
\affiliation{Department of Physics, University of Notre Dame, Notre Dame, Indiana 46556}
\author{D.~\surname{Joffe}}
\affiliation{Department of Physics, Northwestern University, Evanston, Illinois 60208}
\author{I.~\surname{Kachaev}}
\affiliation{Institute for High Energy Physics, Protvino, Russian Federation 142284}
\author{W.~\surname{Kern}}
\affiliation{Department of Physics, University of Massachusetts Dartmouth, North Dartmouth, Massachusetts 02747}
\author{E.~\surname{King}}
\affiliation{Department of Physics, University of Massachusetts Dartmouth, North Dartmouth, Massachusetts 02747}
\author{O.~L.~\surname{Kodolova}}
\affiliation{Nuclear Physics Institute, Moscow State University, Moscow, Russian Federation 119899}
\author{V.~L.~\surname{Korotkikh}}
\affiliation{Nuclear Physics Institute, Moscow State University, Moscow, Russian Federation 119899}
\author{M.~A.~\surname{Kostin}}
\affiliation{Nuclear Physics Institute, Moscow State University, Moscow, Russian Federation 119899}
\author{J.~\surname{Kuhn}}
\affiliation{Department of Physics, Rensselaer Polytechnic Institute, Troy, New York 12180}
\author{V.~V.~\surname{Lipaev}}
\affiliation{Institute for High Energy Physics, Protvino, Russian Federation 142284}
\author{J.~M.~\surname{LoSecco}}
\affiliation{Department of Physics, University of Notre Dame, Notre Dame, Indiana 46556}
\author{M.~\surname{Lu}}
\affiliation{Department of Physics, Rensselaer Polytechnic Institute, Troy, New York 12180}
\author{J.~J.~\surname{Manak}}
\affiliation{Department of Physics, University of Notre Dame, Notre Dame, Indiana 46556}
\author{J.~\surname{Napolitano}}
\affiliation{Department of Physics, Rensselaer Polytechnic Institute, Troy, New York 12180}
\author{C.~\surname{Olchanski}}
\altaffiliation[Present address: ]{TRIUMF, Vancouver, B.C., V6T 2A3, Canada}
\affiliation{Physics Department, Brookhaven National Laboratory, Upton, New York 11973}
\author{A.~I.~\surname{Ostrovidov}}
\altaffiliation[Present address: ]{Department of Physics, Florida State University, Tallahassee, FL 32306}
\affiliation{Nuclear Physics Institute, Moscow State University, Moscow, Russian Federation 119899}
\author{T.~K.~\surname{Pedlar}}
\altaffiliation[Present address: ]{Laboratory for Nuclear Studies, Cornell University, Ithaca, NY 14853}
\affiliation{Department of Physics, Northwestern University, Evanston, Illinois 60208}
\author{A.~V.~\surname{Popov}}
\affiliation{Institute for High Energy Physics, Protvino, Russian Federation 142284}
\author{D.~I.~\surname{Ryabchikov}}
\affiliation{Institute for High Energy Physics, Protvino, Russian Federation 142284}
\author{L.~I.~\surname{Sarycheva}}
\affiliation{Nuclear Physics Institute, Moscow State University, Moscow, Russian Federation 119899}
\author{K.~K.~\surname{Seth}}
\affiliation{Department of Physics, Northwestern University, Evanston, Illinois 60208}
\author{N.~\surname{Shenhav}}
\affiliation{Department of Physics, University of Massachusetts Dartmouth, North Dartmouth, Massachusetts 02747}
\author{X.~\surname{Shen}}
\altaffiliation[Permanent address: ]{Institute of High Energy Physics, Bejing, China}
\affiliation{Thomas Jefferson National Accelerator Facility, Newport News, Virginia 23606}
\author{W.~D.~\surname{Shephard}}
\affiliation{Department of Physics, University of Notre Dame, Notre Dame, Indiana 46556}
\author{N.~B.~\surname{Sinev}}
\affiliation{Nuclear Physics Institute, Moscow State University, Moscow, Russian Federation 119899}
\author{D.~L.~\surname{Stienike}}
\affiliation{Department of Physics, University of Notre Dame, Notre Dame, Indiana 46556}
\author{J.~S.~\surname{Suh}}
\affiliation{Physics Department, Brookhaven National Laboratory, Upton, New York 11973}
\author{S.~A.~\surname{Taegar}}
\affiliation{Department of Physics, University of Notre Dame, Notre Dame, Indiana 46556}
\author{A.~\surname{Tomaradze}}
\affiliation{Department of Physics, Northwestern University, Evanston, Illinois 60208}
\author{I.~N.~\surname{Vardanyan}}
\affiliation{Nuclear Physics Institute, Moscow State University, Moscow, Russian Federation 119899}
\author{D.~P.~\surname{Weygand}}
\affiliation{Thomas Jefferson National Accelerator Facility, Newport News, Virginia 23606}
\author{D.~B.~\surname{White}}
\affiliation{Department of Physics, Rensselaer Polytechnic Institute, Troy, New York 12180}
\author{H.~J.~\surname{Willutzki}}
\altaffiliation{Deceased, Dec. 2001}
\affiliation{Physics Department, Brookhaven National Laboratory, Upton, New York 11973}
\author{M.~\surname{Witkowski}}
\affiliation{Department of Physics, Rensselaer Polytechnic Institute, Troy, New York 12180}
\author{A.~A.~\surname{Yershov}}
\affiliation{Nuclear Physics Institute, Moscow State University, Moscow, Russian Federation 119899}

\collaboration{The Brookhaven E852 Collaboration}
\noaffiliation

\date{June 12, 2002}

\pagebreak
\begin{abstract}
The reaction $\pi^-p \rightarrow \omega\pi^- p$, $\omega \rightarrow \pi^+ \pi^-\pi^0$ has been studied at $18$ GeV/$c$.  The $\omega \pi^-$ mass spectrum is found to be dominated by the $b_1(1235)$.  Partial Wave Analysis (PWA) shows that
$b_1$ production is dominated by natural parity exchange.  The $S$-wave and $D$-wave amplitudes for $b_1(1235) \rightarrow \omega \pi$ have been determined, and it is found that the amplitude ratio, $|D/S| = 0.269\pm ({0.009})_{\text{stat}}\pm ({0.01})_{\text{sys}}$ and the phase difference, $\phi(D - S) = 10.54^\circ \pm ({2.4^\circ})_{\text{stat}}\pm({3.9})_{\text{sys}}$.


\end{abstract}
\maketitle
\section{INTRODUCTION}
According to QCD mesons are bound states of quarks, anti-quarks, and gluons.  The interaction between quarks and gluons is most conveniently simulated by a two part potential.  The short range behavior of the interaction is dominated by the one gluon exchange `Coulombic' potential, while the long range part is dominated by a `linear' confinement potential which is attributed to a collective multi-gluon exchange, often modeled as a flux-tube.  Meson decays, for example, A$\rightarrow$B$+$C, necessarily arise from the confinement interaction and the corresponding breaking of the flux-tube with the creation of a $q\bar{q}$ pair in a certain $^{2s+1}L_J$ state.  Several different calculations exist in the literature~\cite{ly1,ly2,ki,kp,gs,abs,bbs}, assuming a $^3S_1$ pair creation~\cite{ki,kp,gs}, a $^3P_0$ pair creation~\cite{ly1,ly2,ki,gs,abs}, and with  and without final state interactions~\cite{kp,gs,bbs}.  It was pointed out by Kokoski and Isgur~\cite{ki} that accurate measurements of amplitude ratios in specific decays such as the $D$-wave to $S$-wave amplitude ratio, $(D/S)$, in $b_1 \rightarrow \omega \pi$, $a_1 \rightarrow \rho \pi$, and $h_1 \rightarrow \rho \pi$, or the $P$-wave to $F$-wave amplitude ratio, $(P/F)$, in $\pi_2 \rightarrow \rho \pi$, could provide sensitive measures which distinguish between different model predictions.  This is so because these decay amplitude ratios are more sensitive to the decay dynamics than to the hadronic structures~\cite{gs}.  

The experimental measurements compiled by the Particle Data Group (PDG2000)~\cite{epj}, yield a wide range of $|D/S|$ values for $b_1(1235) \rightarrow \omega \pi$ and there are no previous measurements of $\phi(D-S)$.  The PDG2000 recommended $|D/S|$ value of $0.29 \pm 0.04$ is taken as the weighted average of several measurements.  The two most recent measurements, $0.23\pm0.03$ and $0.45\pm0.04$, reported by the Crystal Barrel collaboration from the analyses of the $\omega \eta \pi^0$~\cite{amsler1} and $\omega \pi^0 \pi^0$~\cite{amsler2} final states in $\bar{p}p$ annihilation at rest, are not consistent with each other.  The remaining experimental measurements from $b_1$ production in $\gamma p$~\cite{atkinson} and $\pi p$~\cite{gessaroli,chung,chaloupka,karshon} reactions were based on very low statistics and none were sensitive enough to determine $\phi(D-S)$.

We have made a high statistics measurement of $b_1(1235) \rightarrow \omega \pi$ produced in the reaction $\pi^- p \rightarrow \omega \pi^- p$ at $18$ GeV/$c$, with $\omega \rightarrow \pi^+ \pi^- \pi^0$ and $\pi^0 \rightarrow \gamma \gamma$.  The final state particles $\pi^+,2\pi^-$, and $2\gamma$ were detected; the recoil proton trajectory was also measured.  The $\omega \rightarrow 3\pi$ decay matrix element was used to estimate the signal from the background of other $3\pi$ production processes.  The $J^{pc} = 1^{+-} \, b_1(1235)$ signal was identified by a Partial Wave Analysis (PWA) of the data.  The $b_1 \rightarrow (\omega \pi)_S$ and $b_1 \rightarrow (\omega \pi)_D$ amplitudes were determined, yielding new, accurate measurements of both $|D/S|$ and $\phi(D-S)$.

\section{DATA SELECTION AND MAIN FEATURES OF THE DATA}
The present measurements were part of the Brookhaven National Lab experiment E852, performed at the Multi-Particle Spectrometer (MPS) facility which has been described in detail elsewhere~\cite{three_pi}.  An $18.3$ GeV/$c$ $\pi^-$ beam, delivered by the Alternating Gradient Synchrotron, was incident on a $30$ cm liquid Hydrogen target.  The MPS was equipped with a $4$-layer cylindrical wire chamber (TCYL)~\cite{tcyl} for triggering and detection of charged recoil particles, a $198$-block cylindrical Thallium doped Cesium Iodide detector~\cite{csi} around TCYL to veto soft photons, a $3045$-element lead glass calorimeter (LGD)~\cite{lgd} for detection of photons, and a downstream $2$-plane drift chamber, located directly in front of LGD, for tagging charged particles incident on the LGD.  Three proportional wire chambers (PWCs) were interspersed between six $7$-plane drift chambers~\cite{drift} inside the MPS magnet.  The first two PWCs provided a forward multiplicity trigger.  A multilayer lead/scintillator sandwich in the shape of a picture frame consisting of four box shaped sections was placed after the second PWC.  This detector in conjunction with four scintillators allowed for rejection of downstream wide-angle photons that fell outside the acceptance of the LGD.

The trigger for the reaction $\pi^-p \rightarrow \omega\pi^- p,\; \omega \rightarrow \pi^+\pi^-\pi^0, \;\pi^0\rightarrow \gamma \gamma$  required three forward-going charged particles, as well as one large angle charged recoil (i.e. the final state proton) in TCYL.  A total of $265\,$ million such triggers were recorded during the $1995$ running period of E852.  Photons from $\pi^0 \rightarrow \gamma \gamma$ were detected in the LGD.  After requiring charge and energy-momentum conservation in addition to topological and fiducial cuts, $8.2\,$ million events of the type $\pi^+\pi^-\pi^- \gamma \gamma$ and a missing mass around the mass of the proton remained.  
A 2 constraint kinematical fit, requiring a proton recoil at the main vertex and a $\pi^0$ from the $2\gamma$'s, with a confidence level (c.l.)$> 5\%$, resulted in $1.2\,$ million $\pi^+\pi^-\pi^0\pi^- p$ events.  An additional kinematical constraint requiring the 
$\pi^+\pi^-\pi^0$ mass to be consistent with the $\omega$ mass was imposed.  Events with a c.l. $> 5\%$ were
selected, resulting in a final sample of $224\,$ thousand $\omega\pi^-p$ exclusive events which were then subjected to more detailed analysis. 

\begin{figure}[htbp] 
\includegraphics[width=3.1in]{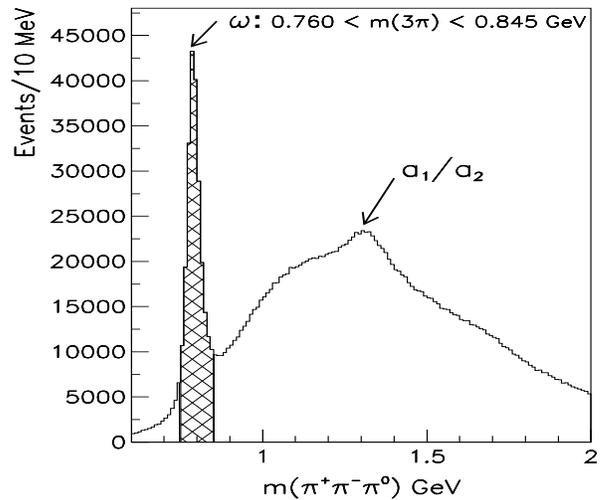}
\caption{Invariant mass spectrum for $\pi^+ \pi^- \pi^0$}
\label{paper_3pi_mass}
\end{figure}

Figure~\ref{paper_3pi_mass} shows the $\pi^+\pi^-\pi^0$ invariant mass spectrum.  There are two entries per event, corresponding to the two neutral three pion combinations.  The hatched region corresponds to a cut around the $\omega$ peak, defined as $0.760<m(\pi^+\pi^-\pi^0)<0.845\,$ GeV/$c^2$, which dominates the spectrum.  There is also a significant number of events in the region of the $a_1(1260)$ and the $a_2(1320)$.  

\begin{figure}[htbp] 
\includegraphics[width=3.1in]{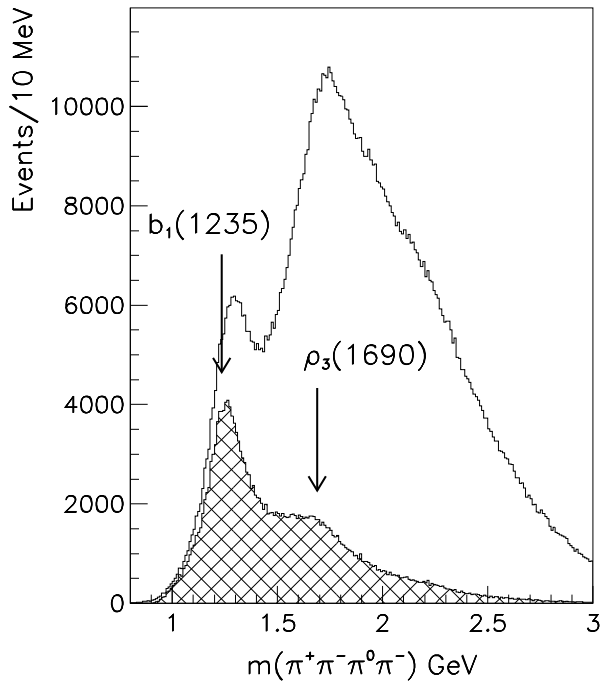}
\caption{Invariant mass spectrum for $\pi^+ \pi^- \pi^0 \pi^-$ and $\omega \pi^-$ combinations.  The hatched regions correspond to an $\omega$ mass cut.}
\label{paper_4pi_mass}
\end{figure}

The $\pi^+\pi^-\pi^0\pi^-$ effective mass spectrum is shown in Figure~\ref{paper_4pi_mass}, before (un-hatched) and after (hatched) the $\omega$ selection.  The $4\pi$ spectrum shows two distinct peaks after the $\omega$ cut, one around the $b_1(1235)$ mass, and one at the $\rho_3(1690)$ mass.  As seen in Figure~\ref{paper_3pi_mass}, there is a significant, approximately linearly increasing background under the $\omega$ peak.  Its magnitude is on the order of $25\%$, and it arises mainly from the $a_1/a_2$ resonances, which also decay to $3\pi$.  This is confirmed from the PWA results discussed in section $6$. 

\begin{figure}[hbt!]
\includegraphics[width=3.in]{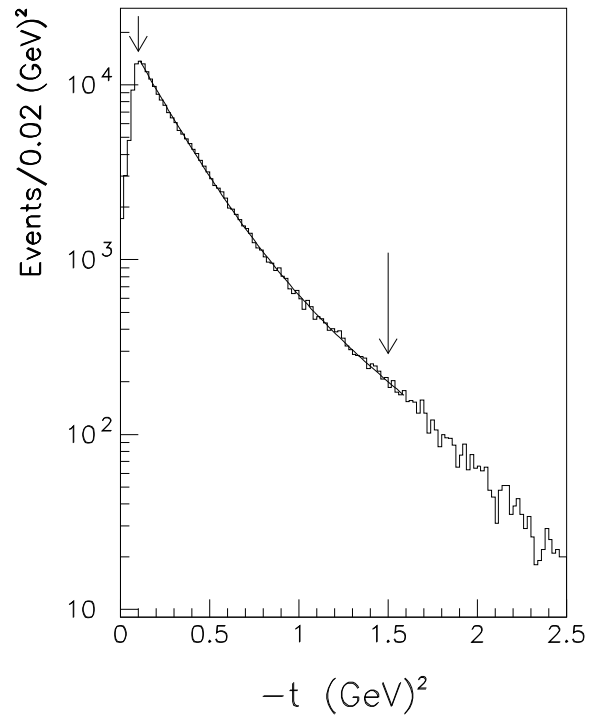}
\caption{The negative of the four-momentum transfer squared, $-t$ distribution for the $\omega\pi^-\,{\rm p}$ final state.  The distribution is fitted to the sum of two exponential functions.  The fit results yield two slopes, $P2=4.5\pm0.052$ (GeV$^2$)$^{-1}$ and $P4=1.7\pm0.070$ (GeV$^2$)$^{-1}$.  The arrows indicate the region chosen in the PWA: ($0.1 <-t<1.5\; {\rm (GeV)}^2$).}
\label{133_om_cuts-paper-t}
\end{figure}

The distribution of four-momentum transfer squared, $-t$, is shown in Figure~\ref{133_om_cuts-paper-t}.  For the PWA only data with $0.1 <-t<1.5\; {\rm (GeV)}^2$ were used.  In this region, the $-t$ distribution was fitted to a function of the form $f(t) = P_1\,e^{P_2\,t} + P_3\,e^{P_4\,t}$, with the coefficients $P_1 = 9.9\pm0.01$ and $P_3 = 7.7\pm0.10$, and the slope values $P_2=4.5\pm0.052$ (GeV$^2$)$^{-1}$ and $P_4=1.7\pm0.070$ (GeV$^2$)$^{-1}$. 
 
\section{PWA OF THE $\omega\pi^-$ SYSTEM}
The details of the E852 PWA formalism are discussed in~\cite{SU,SU2} and a general description on the 
implementation of the formalism, using a PWA code, is given in~\cite{DWJC}.  In the formalism, the interaction process is divided into 
two parts, the production and the decay.  Calculation of the decay amplitudes in our PWA formalism follows the framework of the isobar model~\cite{herndon}, where the decays at each vertex proceed through two-body modes.  An exception to this assumption is the treatment of the $\omega$ decay, where a direct three-body decay is used. 


In the PWA, each possible resonance is characterized by a partial wave, labeled by the 
quantum numbers $J^{pc}[isobar]L {m}^\epsilon$, where $J^{pc}$ are the total spin, parity and 
C-parity of the partial wave, $L$ is the orbital angular momentum between the decay products, $m$ is the absolute value of the spin projection 
of the resonance along the quantization axis (chosen to be in the beam direction), and $\epsilon$ is the reflectivity of the partial wave.  The reflection operator is defined as a rotation about the $y$-axis by $\pi$ radians followed by the parity operation.  In our analysis, the $y$-axis is chosen in the direction of the normal to the production plane. 

The amplitudes, expressed as eigenstates of the reflection operator, are constructed from 
the helicity states, to account for parity conservation in the production~\cite{SULT}.  There 
are two advantages to using the reflectivity basis.  Firstly, the states of different $\epsilon$ do not interfere with 
each other, and secondly, in $\pi {\rm p}$ reactions, there is a direct correlation between $\epsilon$ and the naturality of the exchanged particle.  

The total intensity distribution is written as the sum of $2k$ intensities where each intensity is the square of the sum of 
$N_\epsilon$ interfering amplitudes, and the factor of $2$ is for $\epsilon=+1$ and $\epsilon=-1$.  $k$ corresponds 
to the possible configurations of the spin at the baryon vertex, i.e. spin flip and spin non-flip ($k$ is therefore the 
rank of the spin-density matrix).  $N_\epsilon$ is the number of partial waves included in the fit, in a given reflectivity.  The intensity distribution in terms of production amplitudes, $V$, and decay amplitudes, $A$, is given by:

\begin{eqnarray}
I(\tau) = \sum_{k\epsilon} \left \{ \left|\sum_{\alpha} {^\epsilon V_\alpha} {^\epsilon\!A_\alpha(\tau)} \right|^2 \right \}
\end{eqnarray}

The subscript $\alpha$ denotes a set of parameters specifying the interfering amplitudes, for instance, total spin of the state, $J$, its parity, $p$, the component of the total spin along z, $m$, and the orbital angular momentum between its decay products, $L$. $\tau$ is a set of independent variables which specify the configuration of the final state.  It includes the angles of the decay products and their masses in predefined frames.  The decay amplitudes are calculated for each event and the ``unknown'' production amplitudes are varied to obtain the best match between the predicted and the observed intensity distribution through a maximum likelihood fit.  

The ln(likelihood) function, in its final form, is written as:
\begin{eqnarray}
\ln({\cal L}) &=&\sum_i^n \ln \left[\, \sum_{k\epsilon \alpha\alpha'} {^\epsilon V}_{\alpha k} {^\epsilon V_{\alpha' k}^*} {^\epsilon\!A_{\alpha}}(\tau_i) {^\epsilon\!A_{\alpha'}^*}(\tau_i)\right]  \nonumber\\ 
& &  -n \left[\, \sum_{k\epsilon\alpha\alpha'} {^\epsilon V_{\alpha k}} {^\epsilon V_{\alpha' k}^*}  {^\epsilon\!\Psi_{\alpha\alpha'}^a} \right] 
\label{eq:like}
\end{eqnarray}

The first sum is over the number of events in a given mass bin in which a fit is done. The argument of the ln is just the intensity for each event, $I(\tau_i)$.  The experimental acceptance, determined by Monte Carlo simulation, was incorporated into the PWA as a normalization factor on a wave by wave basis.  Two sets of normalization integrals were calculated, the accepted, $\Psi^a$, and the raw, $\Psi^r$, normalization integrals.  The accepted normalization integrals were defined as:

\begin{equation}
{^\epsilon\!\Psi_{\alpha\alpha'}^a} = \frac{1}{M_a} \sum_{i}^{M_a} {^\epsilon\!A_\alpha(\tau_i)} {^\epsilon\!A_{\alpha'(\tau_i)}^*}  
\end{equation}

Where $M_a$ is the number of accepted Monte Carlo events in a given $\omega \pi$ mass bin.  The raw normalization integrals are defined the same, but calculated from the raw normalization integrals, over the number of raw events in a given $\omega \pi$ mass bin.

With  $\eta = M_a/M_r$, the acceptance corrected intensities are written as:
\begin{equation}
I(\tau) = \frac{n}{\eta} \sum_{k \epsilon \alpha\alpha'} {^\epsilon V}_{\alpha k} {^\epsilon V_{\alpha' k}^*} {^\epsilon\!\Psi_{\alpha\alpha'}^r}
\end{equation}


Numerous fits were performed on the final sample of $168\,$ thousand $\omega \pi^-$ events with various sets of allowed partial waves in the fit, different $\omega \pi^-$ mass bin widths, two different regions in $-t$, different starting values for the fit parameters, and different ranks (rank=1 and rank=2).  The general features of the fits did not change significantly in any case and since the rank=2 fit results did not significantly improve the description of the data, a rank=1 fit proved to be sufficient.

The set of waves included in the final PWA fit consisted of $21$ waves with an $\omega$ in the final state, as shown in Table~\ref{waves}, and $16$ waves with $a_1(1260)$ and $a_2(1320)$ isobars in the final state.  The choice of the wave set with $a_1/a_2$ isobars was based on obtaining an adequate description of the angular distribution of the $\omega$ sideband events~\footnote{ The  $\omega$  sideband regions were defined as $0.655 <m(\pi^+\pi^-\pi^0)< 0.725 \;$ GeV and $0.865<m(\pi^+\pi^-\pi^0)< 0.935$ GeV.}.  The partial wave intensities for waves with $J\ge 4$ were found to be insignificant and were eliminated from the fits at an early stage of the analysis.  A non-interfering ``isotropic'' background wave was included in the fit as a cumulative effect of all the small waves omitted from the fit.  
\begin{table}[htb!]
\caption{Set of 21 waves included in the final PWA fit, with an $\omega$ in the final state.}
\label{waves}
\begin{center}
\begin{minipage}{2.5in}
\begin{ruledtabular}
\begin{tabular}{ccc}
$J^{pc}$ & $m^{\epsilon}$ & $L$  \\  \hline
$1^{--}$ & $0^{-},\,1^{+},\,1^{-}$ & $1$  \\ 
$1^{+-}$ & $0^{+},\,1^{+},\,1^{-}$ & $0$  \\
$1^{+-}$ & $0^{+},\,1^{+},\,1^{-}$ & $2$  \\ 
$2^{--}$ & $0^{+},\,1^{+},\,1^{-}$ & $1$ \\ 
$2^{--}$ & $0^{+},\,1^{+},\,1^{-}$ & $3$  \\ 
$2^{+-}$ & $0^{-},\,1^{+},\,1^{-}$ & $2$  \\ 
$3^{--}$ & $0^{-},\,0^{+},\,0^{-}$ & $3$ 
\end{tabular}
\end{ruledtabular}
\end{minipage}
\end{center}
\end{table}

\begin{figure}[h!]
\includegraphics[width=3.2in]{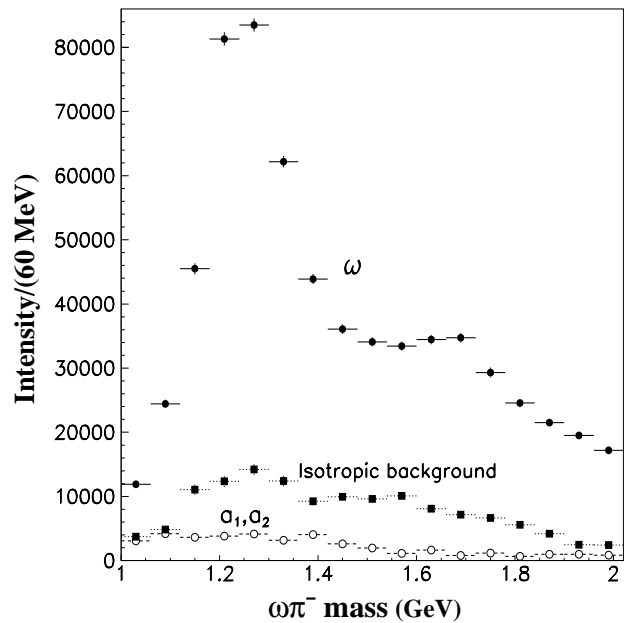}
\caption{Acceptance-corrected intensities from various contributions: Filled circles: waves with an $\omega$ in the final state, Squares: ``isotropic wave, Open circles: waves with either an $a_1$ or an $a_2$ in the final state.}
\label{ompim_fit3-1-paper_2}
\end{figure}

The acceptance-corrected intensities from various contributions are shown in Figure~\ref{ompim_fit3-1-paper_2}.  The three sets of 
points from top to bottom correspond to contributions from waves with an $\omega$ in the final state, a non-interfering isotropic 
background wave, and  waves with either an $a_1$ or an $a_2$ in the final state.  Each point in the plots is the result of an 
independent PWA fit for the events in a $60$ MeV wide mass bin.  As shown in the figure, the total intensity of the $a_1/a_2$ waves was less than $2\%$ of the $\omega$ waves, and for brevity, they are not listed individually.   

\begin{figure}[htb!]
\includegraphics[width=3.4in]{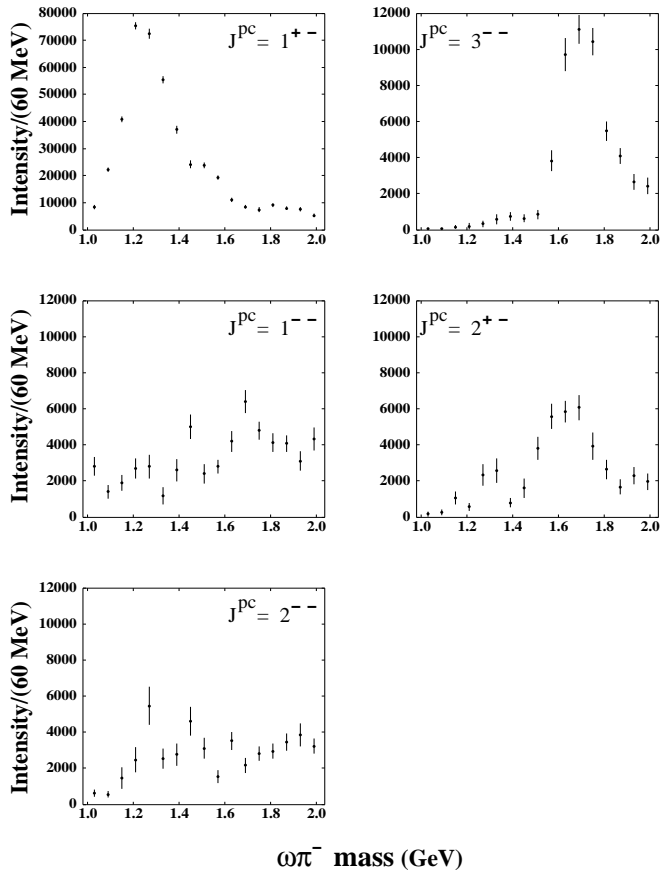}
\caption{Acceptance-corrected intensities from the results of a PWA fit.  The list of $\omega$ waves included in this fit is shown in Table~\ref{waves}.  Individual contributions for different $J^{pc}$ are shown.  Each $J^{pc}$ is the sum of the allowed $m^\epsilon L$ included in the fit.}
\label{ompim_fit3-2-paper_3}
\end{figure}

Figure~\ref{ompim_fit3-2-paper_3} shows the contributions from the different $J^{pc}$ states with an $\omega$ in the final state.  The major contributors to the total intensity are for $J^{pc} = 1^{+-}$ (dominated by $b_1(1235)$), and $J^{pc} = 3^{--}$ (dominated by $\rho_3(1690)$).  We wish to also note that there appears to be a significant enhancement in the $J^{pc} = 2^{+-}$ intensity at $\sim\! 1650$ MeV.  A detailed study of this potentially exotic state will be the subject of a future publication.

\begin{figure}[hbt!]
\includegraphics[width=3.4in]{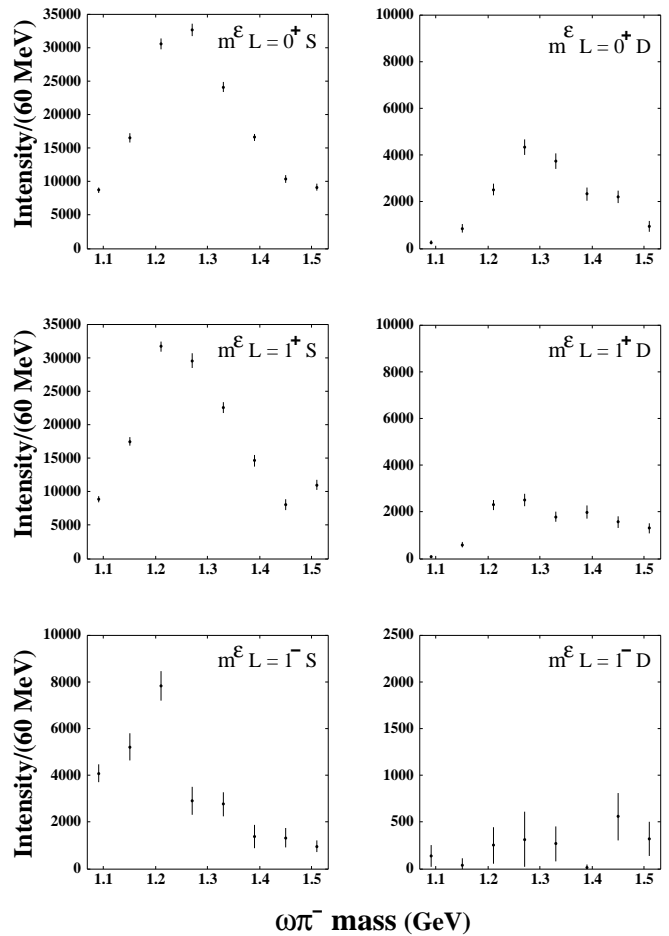}
\caption{Acceptance-corrected intensities for the $J^{pc} = 1^{+-} m^\epsilon L$ partial waves.  Only the positive reflectivity waves are used in the measurement of $D/S$.  Notice the different ordinate scale for the weaker negative reflectivity signals.}
\label{ompim_fit3-3-paper_3}
\end{figure}

\begin{figure}[hbt!]
\includegraphics[width=3.2in]{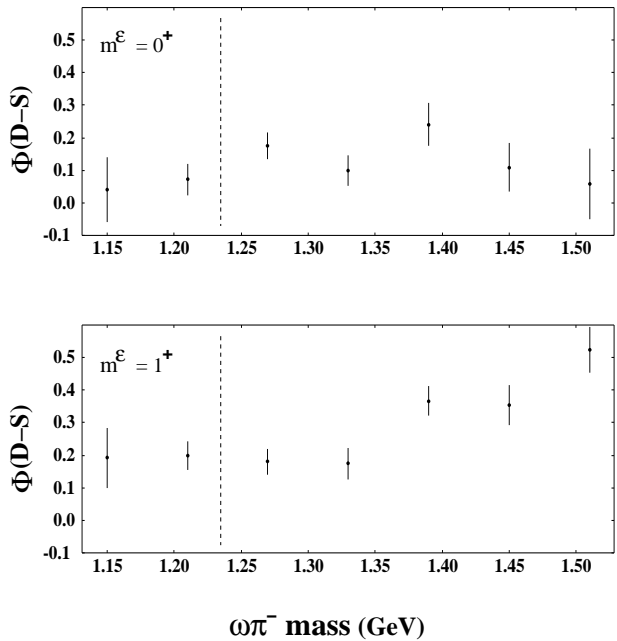}
\caption{Phase differences between the $1^{+-} m^\epsilon$ $D$ and $S$ waves with positive reflectivity, $m^\epsilon = 0^+ \,(1^+)$ in the top (bottom) plot.  The dashed lines are drawn at the mass of the $b_1$ to guide the eyes.}
\label{ompim_fit3-4-paper_3}
\end{figure}

The individual partial wave contributions for $J^{pc} = 1^{+-}$ with different $m^\epsilon L$ are shown
in Figure~\ref{ompim_fit3-3-paper_3} and the corresponding phase differences between the $D$ and $S$ waves of the same $m^\epsilon$ are shown in Figure~\ref{ompim_fit3-4-paper_3}.  It is clear from the intensity plots that $b_1$ production is dominated by the natural parity exchanges.  As expected, the phase difference for the $m^\epsilon = 0^+$ and $1^+$ waves are approximately constant, even over the extended $b_1$ mass region ($\sim\!1.1-1.5$ GeV).  The behavior of the $m^\epsilon = 1^-$ phase difference (not shown) is erratic due to the small intensities and the large error bars.  Only the positive reflectivity waves are used in the measurement of the $D/S$.

\section{$D/S$ MAGNITUDE AND PHASE MEASUREMENT}
From the final set of PWA fits we determine the individual $J^{pc} = 1^{+-}\; {m}^\epsilon$ production amplitudes, where $m^\epsilon=0^+,1^\pm$.  As mentioned earlier, the $-$($+$) reflectivity corresponds to an unnatural(natural) parity exchange.  The $m^\epsilon = 0^+$ and $1^+$ $b_1$ production mechanism is most likely through $\omega$ exchange.  Since the PWA results show that $b_1$ production through the unnatural parity exchanges is small and the error bars on the corresponding negative reflectivity waves is large, they were omitted from the measurement of the $D/S$.  The $D/S$ ratios of both the ${m}^\epsilon = 0^+,1^+$ $b_1$ decay amplitudes were set to a complex number, $R e^{i\phi}$, leaving all other partial waves free to vary independently.  A grid search was then performed in $R$ and $\phi$ for which the -ln(likelihood) function, as written in Eq.~(\ref{eq:like}), was minimized.  Convergence in both $R$ and $\phi$ was reached after few iterations.  The details of the procedure can be found in~\cite{thesis}.

\begin{figure}[htb!]
\includegraphics[width=3.4in]{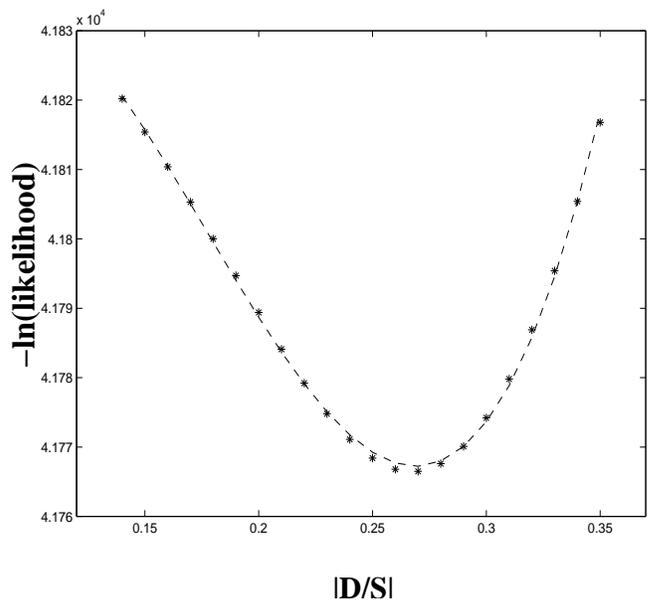}
\caption{The projection of the -ln(likelihood) as a function of $|D/S|$.  The distribution was fitted to a second order polynomial function, with the minimum at  $|D/S| =  0.269\pm 0.009$.}
\label{magscan}
\end{figure}

\begin{figure}[htb!]
\includegraphics[width=3.4in]{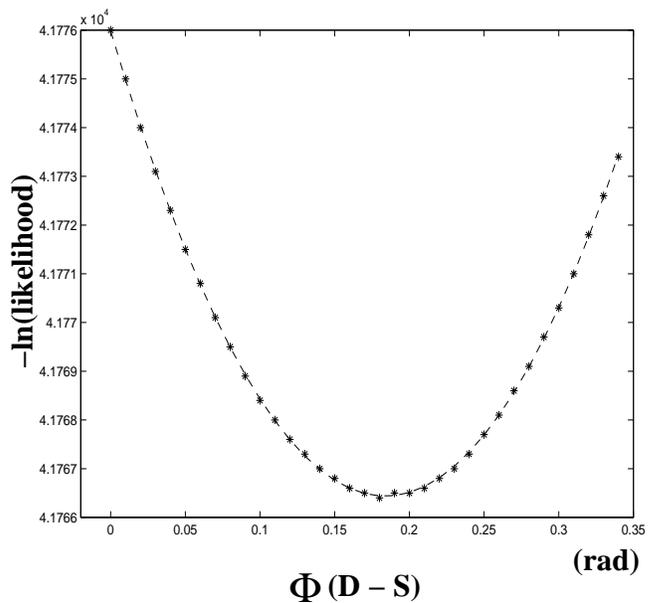}
\caption{The projection of the -ln(likelihood) as a function of $\phi(D-S)$.  The distribution was fitted to a second order polynomial function, with the minimum at $\phi(D-S) = 0.184\pm0.042\,$ rad, or $10.54^\circ\pm2.4^\circ$.} 
\label{phscan}
\end{figure}

The projections of the -ln(likelihood) function at the minimum for $D/S$ magnitude and phase are shown in Figures~\ref{magscan} and ~\ref{phscan} respectively, for the set of waves chosen in the PWA fit as shown in Table~\ref{waves}.  These results are based on a set of $\omega\pi^-$ events in a $160$ MeV wide mass bin around the $b_1$ mass ($1.155-1.315$ GeV) with the $-t$ in the range ($0.1-1.5$ GeV$^2$).  The points in each plot were fit to a second order polynomial function where the minima are found to be $|D/S| = 0.269\pm(0.009)_{\text{stat}}\pm(0.01)_{\text{sys}}$, and $\phi(D-S) = 0.184\pm({0.042})_{\text{stat}}\pm({0.07})_{\text{sys}}$ rad or $10.54 \pm ({2.4})_{\text{stat}}\pm({3.9})_{\text{sys}}$ deg.  The statistical error in each measurement corresponds to the change in \linebreak[4]-ln(likelihood) by $0.5$ units.  The main sources of the systematic error considered were the choice of the wave set in the PWA fits and the size of the $\omega\pi^-$ mass bins used in the scans.  Regarding the choice of the wave sets, two sets of waves with reasonable fit results were used. One set consisted of the waves listed in Table~\ref{waves} and another consisted of a subset of $28$ largest waves from that list.  For the $\omega\pi^-$ mass widths, six sets of independent scans were performed in $60$, $80$, $100$, $120$, $140$ and $160$ MeV wide $\omega \pi$ mass bins around the nominal $b_1$ mass.  No significant systematic change in either $|D/S|$ or $\phi(D-S)$ was observed within the statistical errors, and the systematic errors quoted are a conservative estimate.

Various studies were performed in order to determine the leakage from other waves into all the positive reflectivity $J^{pc} = 1^{+-}$ waves.  In addition, leakage from the $J^{pc} = 1^{+-}$, $S$-wave to $D$-wave for a given $m^\epsilon = 1^-$ was also investigated.  In all cases, the leakage due to experimental resolutions was determined to be negligible. 

\section{CONCLUSIONS}
We have made a study of the reaction $\pi^- p \rightarrow \omega \pi^-$, $\omega \rightarrow \pi^+ \pi^- \pi^0$ at $18$ GeV/$c$.  A partial wave analysis of $168$ thousand events, consistent with the $\omega \pi^- p$ hypothesis, shows that the $\omega \pi^-$ data below $1.6$ GeV are dominated by the $J^{pc} = 1^{+-}$ resonance $b_1(1235)$.  The $S$-wave and $D$-wave amplitudes of the $b_1 \rightarrow \omega \pi^-$ decay have been determined.  It is found that the ratio of the amplitudes, $|D/S| = 0.269\pm(0.009)_{\text{stat}}\pm(0.01)_{\text{sys}}$.  This represents nearly a factor three improvement in error over the current PDG2000 average value of $0.29 \pm0.04$.  We have also determined the phase difference, $\phi(D-S) = 0.184\pm({0.042})_{\text{stat}}\pm({0.07})_{\text{sys}}$ radians, or $10.54 \pm ({2.4})_{\text{stat}}\pm({3.9})_{\text{sys}}$ deg, for which no prior results exist.  Figure~\ref{phscan} shows that the minimum value of $\phi(D-S)$ allowed by the above errors differs from zero by more than $15\sigma$.  The level of significance varies in other acceptable fits, but it is greater than $5\sigma$ in all cases.

We can compare our results with the theoretical predictions which exist in the literature.  Ackleh, Barnes and Swanson~\cite{abs} have made numerical predictions for $|D/S|$ ratios in their $^3P_0$ hadronic decay model as a function of the oscillator parameter $\beta$.  For the then current values of $|D/S|$ for $b_1 \rightarrow \omega \pi$ and $a_1 \rightarrow \rho\pi$, they found the best fit value of $\beta = 0.448$ GeV and the corresponding $|D/S| = 0.219$.  They note, however, that these results are in disagreement with ``the decay rates of light $L(q\bar{q}) = 0$ and $L(q\bar{q}) = 1$ mesons (which) support $\beta = 0.40$ GeV''.  We note that our result, $|D/S| = 0.269\pm 0.013$ corresponds to $\beta = 0.409 \pm 0.008$, in excellent agreement with that for $L=0$ and $L=1$ light meson decays.

Ackleh {\it et al.} also make the parameter independent prediction that the ratio of ratios, \linebreak $R = [(D/S)_{a_1\rightarrow \rho \pi}] / [(D/S)_{ b_1 \rightarrow \omega \pi}] = -0.5$.  Using the current PDG2000 average, $[(D/S)_{a_1\rightarrow \rho \pi}] = -0.107 \pm 0.016$ with our result for $(D/S)$ for $b_1 \rightarrow \omega \pi$, we obtain $R = -0.40 \pm 0.06$, with the error determined almost entirely by that in the $a_1$ decay ratio.  This points at the need to improve the measurement of $(D/S)$ for $a_1\rightarrow \rho\pi$, which is plagued by uncertainty in the Deck contribution to this decay.

The phase difference between the $D$ and $S$ wave decays of $b_1 \rightarrow \omega \pi$ can arise from the final state interaction between the $\omega$ and $\pi$.  In a quark interchange model calculation, Barnes, Black, and Swanson~\cite{bbs} predict $\phi(D-S) = 14$ deg.  Our measurement, $\phi(D-S) = 10.54\pm ({2.4})_{\text{stat}}\pm({3.9})_{\text{sys}}$ deg, is consistent with their prediction.

\begin{acknowledgments}
We are grateful to the members of the Brookhaven MPS group for their outstanding support in running this experiment.  We wish to dedicate this paper to the memory of our recently deceased colleague, H. J. Willutzki.  This research was supported in part by the US Department of Energy, the US National Science Foundation, and the Russian Ministry of Industry and Science.
\end{acknowledgments}

\bibliography{b1_ds}

\begin{thebibliography}{26}
\expandafter\ifx\csname natexlab\endcsname\relax\def\natexlab#1{#1}\fi
\expandafter\ifx\csname bibnamefont\endcsname\relax
  \def\bibnamefont#1{#1}\fi
\expandafter\ifx\csname bibfnamefont\endcsname\relax
  \def\bibfnamefont#1{#1}\fi
\expandafter\ifx\csname citenamefont\endcsname\relax
  \def\citenamefont#1{#1}\fi
\expandafter\ifx\csname url\endcsname\relax
  \def\url#1{\texttt{#1}}\fi
\expandafter\ifx\csname urlprefix\endcsname\relax\def\urlprefix{URL }\fi
\providecommand{\bibinfo}[2]{#2}
\providecommand{\eprint}[2][]{\url{#2}}

\bibitem[{\citenamefont{{A. Le Yaouanc et al.}}(1973)}]{ly1}
\bibinfo{author}{\bibnamefont{{A. Le Yaouanc et al.}}},
  \bibinfo{journal}{Physical Review D} \textbf{\bibinfo{volume}{8}},
  \bibinfo{pages}{2223} (\bibinfo{year}{1973}).

\bibitem[{\citenamefont{{A. Le Yaouanc et al.}}(1977)}]{ly2}
\bibinfo{author}{\bibnamefont{{A. Le Yaouanc et al.}}},
  \bibinfo{journal}{Physics Letters B} \textbf{\bibinfo{volume}{71}},
  \bibinfo{pages}{397} (\bibinfo{year}{1977}).

\bibitem[{\citenamefont{Kokoski and Isgur}(1987)}]{ki}
\bibinfo{author}{\bibfnamefont{R.}~\bibnamefont{Kokoski}} \bibnamefont{and}
  \bibinfo{author}{\bibfnamefont{N.}~\bibnamefont{Isgur}},
  \bibinfo{journal}{Physical Review D} \textbf{\bibinfo{volume}{35}},
  \bibinfo{pages}{907} (\bibinfo{year}{1987}).

\bibitem[{\citenamefont{Kumano and Pandharipande}(1988)}]{kp}
\bibinfo{author}{\bibfnamefont{S.}~\bibnamefont{Kumano}} \bibnamefont{and}
  \bibinfo{author}{\bibfnamefont{V.}~\bibnamefont{Pandharipande}},
  \bibinfo{journal}{Physical Review D} \textbf{\bibinfo{volume}{38}},
  \bibinfo{pages}{146} (\bibinfo{year}{1988}).

\bibitem[{\citenamefont{Geiger and Swanson}(1994)}]{gs}
\bibinfo{author}{\bibfnamefont{P.}~\bibnamefont{Geiger}} \bibnamefont{and}
  \bibinfo{author}{\bibfnamefont{E.}~\bibnamefont{Swanson}},
  \bibinfo{journal}{Physical Review D} \textbf{\bibinfo{volume}{50}},
  \bibinfo{pages}{6855} (\bibinfo{year}{1994}).

\bibitem[{\citenamefont{Ackleh et~al.}(1996)\citenamefont{Ackleh, Barnes, and
  Swanson}}]{abs}
\bibinfo{author}{\bibfnamefont{E.~S.} \bibnamefont{Ackleh}},
  \bibinfo{author}{\bibfnamefont{T.}~\bibnamefont{Barnes}}, \bibnamefont{and}
  \bibinfo{author}{\bibfnamefont{E.}~\bibnamefont{Swanson}},
  \bibinfo{journal}{Physical Review D} \textbf{\bibinfo{volume}{54}},
  \bibinfo{pages}{6811} (\bibinfo{year}{1996}).

\bibitem[{\citenamefont{Barnes et~al.}(2001)\citenamefont{Barnes, Black, and
  Swanson}}]{bbs}
\bibinfo{author}{\bibfnamefont{T.}~\bibnamefont{Barnes}},
  \bibinfo{author}{\bibfnamefont{N.}~\bibnamefont{Black}}, \bibnamefont{and}
  \bibinfo{author}{\bibfnamefont{E.}~\bibnamefont{Swanson}},
  \bibinfo{journal}{Physical Review C} \textbf{\bibinfo{volume}{63}},
  \bibinfo{pages}{025204} (\bibinfo{year}{2001}).

\bibitem[{\citenamefont{{D.E. Groom et al.}}(2000)}]{epj}
\bibinfo{author}{\bibnamefont{{D.E. Groom et al.}}}, \bibinfo{journal}{TEPJ}
  \textbf{\bibinfo{volume}{15}}, \bibinfo{pages}{425} (\bibinfo{year}{2000}).

\bibitem[{\citenamefont{{C. Amsler et al.}}(1994)}]{amsler1}
\bibinfo{author}{\bibnamefont{{C. Amsler et al.}}}, \bibinfo{journal}{Physics
  Letters B} \textbf{\bibinfo{volume}{327}}, \bibinfo{pages}{425}
  (\bibinfo{year}{1994}).

\bibitem[{\citenamefont{{C. Amsler et al.}}(1993)}]{amsler2}
\bibinfo{author}{\bibnamefont{{C. Amsler et al.}}}, \bibinfo{journal}{Physics
  Letters B} \textbf{\bibinfo{volume}{311}}, \bibinfo{pages}{362}
  (\bibinfo{year}{1993}).

\bibitem[{\citenamefont{{M. Atkinson et al.}}(1984)}]{atkinson}
\bibinfo{author}{\bibnamefont{{M. Atkinson et al.}}}, \bibinfo{journal}{Nuclear
  Physics B} \textbf{\bibinfo{volume}{243}}, \bibinfo{pages}{1}
  (\bibinfo{year}{1984}).

\bibitem[{\citenamefont{{R. Gessaroli et al.}}(1977)}]{gessaroli}
\bibinfo{author}{\bibnamefont{{R. Gessaroli et al.}}},
  \bibinfo{journal}{Nuclear Physics B} \textbf{\bibinfo{volume}{126}},
  \bibinfo{pages}{382} (\bibinfo{year}{1977}).

\bibitem[{\citenamefont{{S. U. Chung et al.}}(1975)}]{chung}
\bibinfo{author}{\bibnamefont{{S. U. Chung et al.}}},
  \bibinfo{journal}{Physical Review D} \textbf{\bibinfo{volume}{11}},
  \bibinfo{pages}{2426} (\bibinfo{year}{1975}).

\bibitem[{\citenamefont{{V. Chaloupka et al.}}(1974)}]{chaloupka}
\bibinfo{author}{\bibnamefont{{V. Chaloupka et al.}}},
  \bibinfo{journal}{Physics Letters B} \textbf{\bibinfo{volume}{51}},
  \bibinfo{pages}{407} (\bibinfo{year}{1974}).

\bibitem[{\citenamefont{{U. Karshon et al.}}(1974)}]{karshon}
\bibinfo{author}{\bibnamefont{{U. Karshon et al.}}}, \bibinfo{journal}{Physical
  Review D} \textbf{\bibinfo{volume}{10}}, \bibinfo{pages}{3608}
  (\bibinfo{year}{1974}).

\bibitem[{\citenamefont{{S.U. Chung et al.}}(2001)}]{three_pi}
\bibinfo{author}{\bibnamefont{{S.U. Chung et al.}}}, \bibinfo{journal}{Physical
  Review D} \textbf{\bibinfo{volume}{60}}, \bibinfo{pages}{092001}
  (\bibinfo{year}{2001}).

\bibitem[{\citenamefont{{Z. Bar$-$Yam et al.}}(1997)}]{tcyl}
\bibinfo{author}{\bibnamefont{{Z. Bar$-$Yam et al.}}},
  \bibinfo{journal}{Nuclear Instruments and Methods A}
  \textbf{\bibinfo{volume}{386}}, \bibinfo{pages}{235} (\bibinfo{year}{1997}).

\bibitem[{\citenamefont{{T. Adams et al.}}(1996)}]{csi}
\bibinfo{author}{\bibnamefont{{T. Adams et al.}}}, \bibinfo{journal}{Nuclear
  Instruments and Methods A} \textbf{\bibinfo{volume}{368}},
  \bibinfo{pages}{617} (\bibinfo{year}{1996}).

\bibitem[{\citenamefont{{R. R. Crittenden et al.}}(1997)}]{lgd}
\bibinfo{author}{\bibnamefont{{R. R. Crittenden et al.}}},
  \bibinfo{journal}{Nuclear Instruments and Methods A}
  \textbf{\bibinfo{volume}{387}}, \bibinfo{pages}{377} (\bibinfo{year}{1997}).

\bibitem[{\citenamefont{{S. E. Eiseman et al.}}(1983)}]{drift}
\bibinfo{author}{\bibnamefont{{S. E. Eiseman et al.}}},
  \bibinfo{journal}{Nuclear Instruments and Methods}
  \textbf{\bibinfo{volume}{217}}, \bibinfo{pages}{140} (\bibinfo{year}{1983}).

\bibitem[{\citenamefont{Chung}(1971)}]{SU}
\bibinfo{author}{\bibfnamefont{S.~U.} \bibnamefont{Chung}}, \bibinfo{type}{CERN
  report 71-08. Lectures given in the Academic Training Program of CERN
  1969-1970.}, \bibinfo{institution}{CERN} (\bibinfo{year}{1971}).

\bibitem[{\citenamefont{Chung}(1995)}]{SU2}
\bibinfo{author}{\bibfnamefont{S.~U.} \bibnamefont{Chung}}, \bibinfo{type}{BNL
  report QGS-93-05.}, \bibinfo{institution}{BNL} (\bibinfo{year}{1995}).

\bibitem[{\citenamefont{Cummings and Weygand}(1997)}]{DWJC}
\bibinfo{author}{\bibfnamefont{J.~P.} \bibnamefont{Cummings}} \bibnamefont{and}
  \bibinfo{author}{\bibfnamefont{D.~P.} \bibnamefont{Weygand}},
  \bibinfo{type}{BNL report 64637.}, \bibinfo{institution}{BNL}
  (\bibinfo{year}{1997}).

\bibitem[{\citenamefont{{D.J. Herndon et al.}}(1975)}]{herndon}
\bibinfo{author}{\bibnamefont{{D.J. Herndon et al.}}},
  \bibinfo{journal}{Physical Review D} \textbf{\bibinfo{volume}{11}},
  \bibinfo{pages}{3165} (\bibinfo{year}{1975}).

\bibitem[{\citenamefont{{S. U. Chung and T. L. Trueman}}(1975)}]{SULT}
\bibinfo{author}{\bibnamefont{{S. U. Chung and T. L. Trueman}}},
  \bibinfo{journal}{Physical Review D} \textbf{\bibinfo{volume}{11}},
  \bibinfo{pages}{633} (\bibinfo{year}{1975}).

\bibitem[{\citenamefont{Nozar}(2002)}]{thesis}
\bibinfo{author}{\bibfnamefont{M.}~\bibnamefont{Nozar}}, Ph.D. thesis,
  \bibinfo{school}{Rensselaer Polytechnic Institute}, \bibinfo{address}{Troy,
  NY} (\bibinfo{year}{2002}).

\end{thebibliography}

\end{document}